\documentclass[%
reprint,
 amsmath,amssymb,
 aps, prl,
floatfix,
]{revtex4-2}

\usepackage{url} 
\usepackage{amsmath,amsfonts,amssymb,mathtools}
\usepackage{mathrsfs} 

\usepackage{siunitx}
\usepackage[normalem]{ulem}

\usepackage{xcolor}
\definecolor{granata}{HTML}{831d1c}
\definecolor{kulblue}{HTML}{116E8A}
\usepackage{caption}
\usepackage[font={color=black,it},figurename=Fig.,labelfont={it}]{caption}

 \newcommand{\bfE}{\mathbf{E}}

\newcommand{\bfJ}{\mathbf{J}}

\newcommand{\bfv}{\mathbf{v}}

\DeclareMathAlphabet\mathbfcal{OMS}{cmsy}{b}{n}

\newcommand{\enE}{\mathscr E}
\newcommand{\re}{$R_E \,$}

\usepackage{capt-of}

\usepackage{notes2bib}

\newcommand{\gl}[1]{{\color{black}#1}}   
\newcommand{\jb}[1]{{\color{black}#1}}   
\newcommand{\rw}[1]{{\color{black}#1}}   

\bibnotesetup{ note-name = , use-sort-key = false
}

\begin{document}

\title{Turbulent energization of electron power law tails during magnetic reconnection}

\author{Giovanni Lapenta}
\affiliation{Department of Mathematics, KU Leuven, Leuven, Belgium}
\affiliation{Space Science Institute, Boulder, Colorado, USA}
\author{Jean Berchem}
\author{Mostafa El Alaoui}
\affiliation{Department of Physics and Astronomy, University of California, Los Angeles, CA, USA}
\author{Raymond Walker}
\affiliation{Department of Earth, Planetary, and Space Sciences, University of California, Los Angeles, CA, USA}

\date{\today}

\begin{abstract}
Earth's magnetotail is an excellent laboratory to study the interplay of reconnection and turbulence in determining electron energization. The process of formation of a power law tail during turbulent reconnection is a documented fact still in need of a comprehensive explanation. 
We conduct a massively parallel particle in cell 3D simulation and use enhanced statistical resolution of the high energy range of the particle velocities to study how
reconnection creates the conditions for the tail to be formed. The process is not direct acceleration by the coherent, laminar, reconnection-generated electric field. Rather, reconnection causes turbulent outflows where energy exchange is dominated by a highly non-gaussian distribution of fluctuations. Electron energization is diffuse throughout the entire reconnection outflow but it is heightened by regions of intensified magnetic field  such as dipolarization fronts traveling towards Earth.
\end{abstract}

\maketitle
\newpage

\jb{A ubiquitous feature} of space plasmas still defying  \jb{comprehension} is the presence of power-law distributed tails of high energy particles \citep{zouganelis2008measuring}. Power law tails \rw{occur} in \jb{heliospheric} plasmas and cosmic rays but we focus here on the \jb{magnetosphere} surrounding  Earth. \rw{Observations from the the Magnetospheric MultiScale (MMS) mission \citep{burch2016electron}, like some of its predecessors, show} three concurrent features: the presence of reconnection-generated high speed flows, the presence of turbulence within these flows and the presence of a population of high energy electrons within these turbulent flows\cite{oieroset2002evidence}. What is the overall mechanism interlinking these three observed processes? Previous studies have already highlighted the link between reconnection outflows and turbulence: \citet{eastwood2009observations} reported   greatly heightened electric and magnetic fluctuation spectra with a power law distribution changing slope at hybrid scales that is characteristic of turbulence. \rw{Three dimensional reconnection simulations show this correctly} \citep{pucci2017properties,pucci2018generation}.  Similarly, \rw{ there is }evidence of a link between  particle acceleration and reconnection \citep{PhysRevLett.89.195001}. Turbulence is one of the oldest mechanisms proposed for particle energization\citep{fermi1949origin}.
Significant work has been done in the past few years regarding
electron energization in magnetic reconnection \citep{dahlin2014mechanisms,dahlin2015electron,dahlin2017role,zhou2018suprathermal} and in turbulence
\citep{zhdankin2019electron,zhdankin2020kinetic,lazarian20193d}. Recently, \citet{comisso2019interplay} investigated the relationship of the two processes in the case of magnetically dominated relativistic plasmas. For the case of the classical plasmas of the heliosphere, the question is still open and of particular relevance to ongoing missions that are observing high energy particles in the magnetosphere \citep{torbert2018electron} and in the solar wind \citep{chhiber2019contextual,mueller2013solar}.

To answer this question, we use a combined approach where we first conduct a global magnetohydrodynamic (MHD) magnetosphere simulation to provide the global forcing for reconnection as well as the general environment where the reconnection outflow jet forms and interacts with its surroundings. When reconnection is already active, we  select the region of interest encompassing the whole outflow jet and launch a fully kinetic particle in cell (PIC) simulation that can study with accuracy the electron and ion motion, resolving directly the mechanisms of particle acceleration. Previous work \citep{ashour2015multiscale,lapenta2017origin} shows the ability of PIC to introduce  the correct physics of reconnection when started from a MHD state that misses important physics. After a short transient, the electrons become decoupled from the ions and produce the typical signature of kinetic reconnection \citep{ashour2015multiscale,ashour2016identifying}, forming much faster jets that carry  substantial energy \citep{lapenta2016multiscale,lapenta2020multiscale}. Also a number of instabilities due to the kinetic physics develops \citep{walker2019embedding}. We describe details of our approach  \rw{in }\citet{walker2019embedding}.

The PIC method has limited ability to resolve the particle energy distribution. Most PIC methods apply the Box–M\"uller  algorithm \cite{box1958note}  to generate a drifting Maxwellian distribution with particles of equal weight. In this approach, very few particles are generated in the tail at high speed and the vast majority have speeds close to the drift. This statistical limitation leads to the inability of PIC to capture phenomena acting primarily on the tail. \citet{byers1970noise} (see also Ch. 17 in \citet{birdsall-langdon})  showed that a different loading biased towards more higher energy particles, but with correspondingly reduced statistical weight, leads to a much improved description of the interaction of waves with the tail of the distribution (e.g. Landau damping).

To remove this limitation, we use two species each for electrons and ions. One species is the core bulk population of thermal particles, and the second species, with reduced statistical weight, represents the power law tail. 
This approach differs from a test particle approach \citep{zhou2018suprathermal} in that even the smaller weight particles still contribute to the overall energy budget. This is an important difference because our treatment still conserves energy and the energy needed to accelerate the particles \rw{comes} from the fields. 
As an alternative, the importance sampling method \citep{rubinstein2016simulation,byers1970noise} can be used to generate uniformly distributed particles in velocity space with weights corrected to represent a Maxwellian: this approach leads to results consistent with those reported here \bibnote[sm1]{See Supplemental Material 1 at [LINK] for a comparison of the results obtained with 2 particle population per species versus importance sampling.}.

In this letter, using this approach, we reach two conclusions. First, the core population receives modest energization but the energy of the power law electrons increases dramatically, explaining the observed prevalence of high energy particles in reconnection outflows. Second, the mechanism of particle acceleration for the high speed tail \jb {is not directly linked} with reconnection. \jb{ Instead, the acceleration comes} from the interaction of the particles with  the high-speed reconnection outflow that forms fronts of increased  magnetic field (called dipolarization  fronts when traveling earthward) and \jb{ leads to the production of} a turbulent cascade. We observe a net energization of the particles, especially in the regions of fastest flows and largest vertical field intensification.

We start our study with a generic global MHD model of Earth in GSM coordinates that assumes a constant solar wind flow with \rw{an} earthward speed $V_x=-530$ \si{km/s}, a density $n=6$ \si{\per\cubic\centi\metre},  a \jb{thermal} pressure $p=180$  \si{pPa} \jb{and a southward directed interplanetary }field $B_z=-8$ \si{nT}. We observe reconnection developing on the nightside of the magnetosphere, in the magnetotail \bibnote[sm2]{See Supplemental Material 2 at [LINK] for a movie of the 3D volume rendering of the ion current at the end  of the simulation (cycle 8000) with  magnetic field lines selected progressively closer to the viewer.}. We consider this reconnecting tail as the initial condition for a PIC  simulation that \rw{includes} the subdomain: $x/R_E=[-38.2, -7]$, $y/R_E=[-7.8, 7.8]$ and $z/R_E=[-10.4, 10.4]$  using the MHD simulation results for initial and boundary conditions.  The PIC simulation uses 400x200x280 cells with 125 particles per cell per species, arranged in a topology of 40x10x14 processors. The time step is $\omega_{pi}\Delta t =0.5$, where the reference density used for computing the plasma frequency is $n_0=0.25$\si{\per\cubic\centi\metre}.  The size of Earth is \gl{$R_E/d_{p0}=14.0$} and the duration of the simulation, 8000 cycles, corresponds to about 30 \si{s} of real time. 

We derive the initial state of the PIC simulation from an MHD state by making the assumption that the distribution is Maxwellian with the density from MHD and species velocities determined from the bulk speed and the MHD current density.  We assume the MHD temperature  is that of the ions, while the electron temperature is 5 times smaller  as is typical in the magnetotail. 

\jb {As mentioned above, we use} two main species (0 for the electrons and 1 for the ions) \jb {for the core bulk population of thermal particles} and capture the low probability tail distribution by using \jb{two} additional  species (2 for the electrons and 3 for the ions)  distributed initially \jb{with} a power law tail. We achieve this by sampling the tail of a kappa distribution \cite{pierrard2010kappa}, with $\kappa=2$, using the modified Box–M\"uller \cite{box1958note} transform algorithm  described in \cite{abdul2014method}, with the  condition  $v>v_{th,s}$. The statistical weight of the tail particles was $10^{-3}$ of the weight of the bulk species to extend the dynamical range by three orders of magnitude. This is an arbitrary choice but using a more neutral importance sampling method does not lead to different conclusions, as shown in the Supplemental Material~\bibnotemark[sm1].

\begin{figure}[htp]
\centering \includegraphics[width=.8\columnwidth]{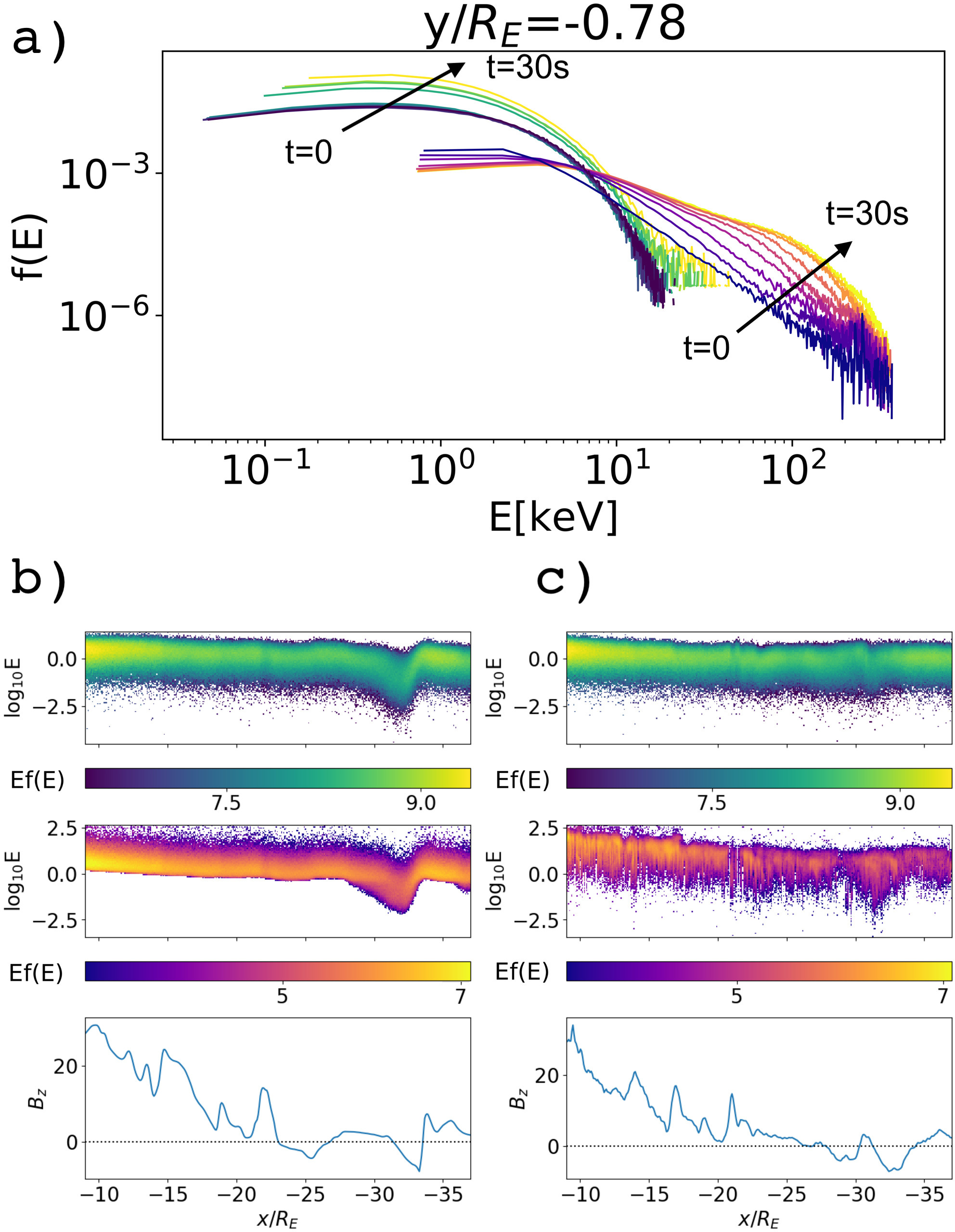}
\caption{\label{eflux} Temporal evolution of the energy distribution. Panel a shows the evolution of the particle distribution function for species 0 (bulk electrons) and species 2 (tail electrons) during the evolution from time 0 to time 30s, in 9 equally spaced linear steps in time (darker to lighter). Panel\rw{s} b and  c report the energy fluxes  $Ef(E)$ of species 0 and species 2 stacked with the vertical magnetic field  $B_z$, at the initial and final times. The distribution is taken for a narrow band \gl{ $y =Ly/2\pm d_i/2$ and $z =Lz/2\pm d_i/2$}. }
\end{figure}


The top panel of Fig.~\ref{eflux} presents evolution of the energy fluxes for the electron species starting at the beginning for the PIC simulation.  The primary conclusion is that the power l\rw{a}w tail electrons are indeed energized much more than the bulk electrons. In particular, Fig.~\ref{eflux}-a shows the electron energy distribution for both species of electrons at the beginning and at different times during the simulation. As can be observed, a power law tail forms extending to high energies. The bulk electrons do not change their distribution by as much. This is consistent with the results from PIC simulations without the high energy tail.

Figure \ref{eflux} (panels b and c)  shows the change in the energy fluxes of electrons in species 0 (bulk) and 2 (power law tail) from the initial to the final time. There is a striking difference between bulk and kappa tail electrons. For the bulk Maxwellian electrons, energization is \gl{present but weak}. For the tail kappa-distributed electrons, energization is by a full order of magnitude everywhere, but larger in the near-Earth region. \gl{By the end of the simulation, energization is starting to saturate.} Attentive inspection reveals  \jb{that the higher energy particles have} a tendency to concentrate in correspondence with the intensification of the vertical component of the magnetic field $B_z$. 

 It is worth noting that in an identical simulation without the additional tail species, no significant power law tails \rw{formed}. Conversely, using importance sampling of a Maxwellian plasma with no initial power law tail leads to substantially similar acceleration of the high energy tail to that reported in Fig.~\ref{eflux} \bibnotemark[sm1]: there is no need to seed a power law initially to obtain one as part of the energization process.

This is our central result: the electric fields generated in the reconnection process do not affect significantly the energy spectrum at low energies but greatly increase  the high energy tail as is well known from observations \jb{(e.g.\cite{oieroset2002evidence})}. The novelty here is the ability to replicate the process in silico, a fact that allows us to investigate correlations between quantities in space and time in a manner not possible from spacecraft observations. 

\begin{figure}[htp]
\centering
\includegraphics[width=\columnwidth]{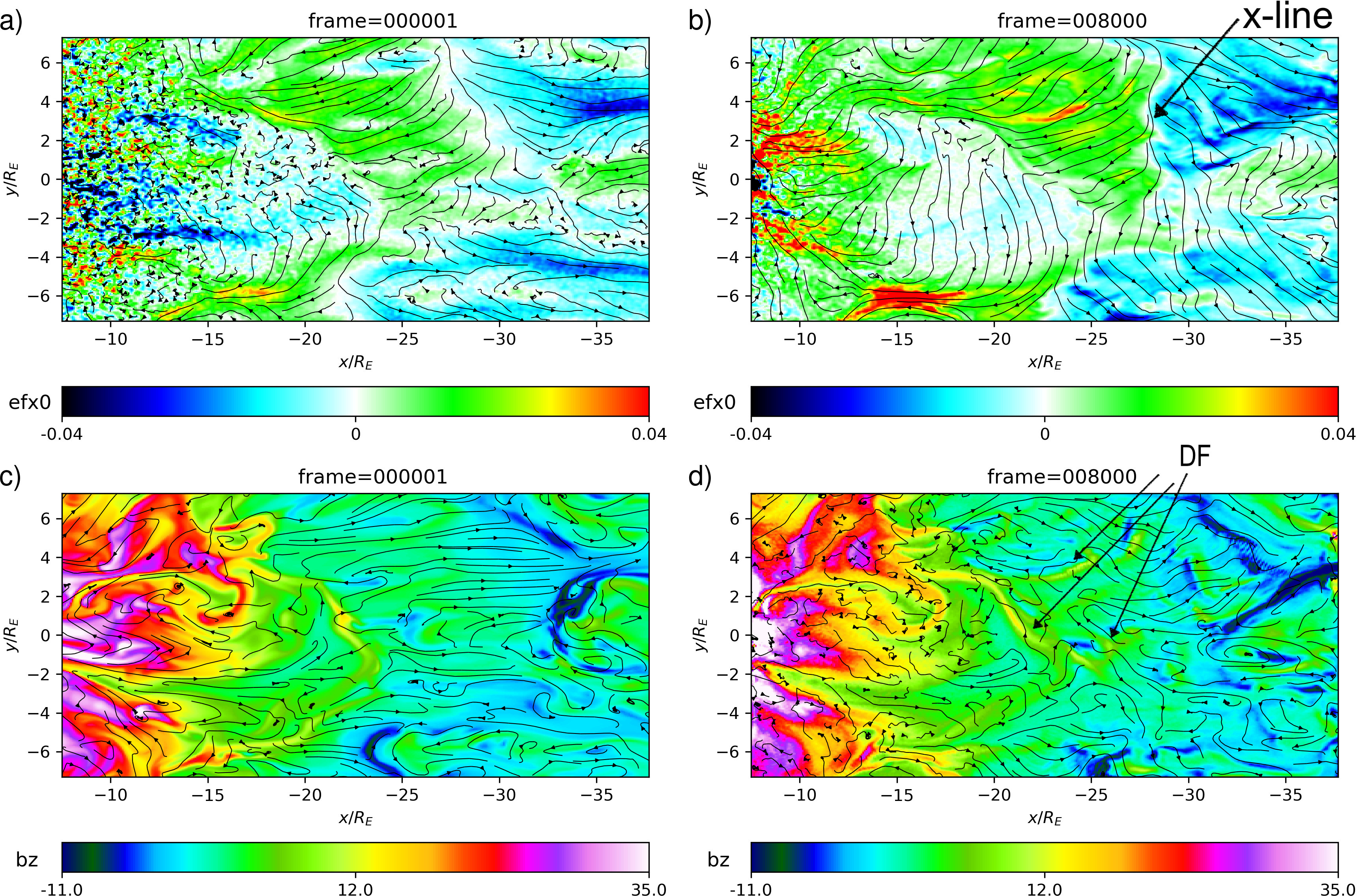}
\caption{ \label{evolution} Temporal evolution of the bulk electron energy flux (top, panels a and b) and of the vertical component of the magnetic field (bottom, panels c and d). Black lines show in the top panels the bulk Maxwellian electron speed (species 0) on the equatorial plane and in the bottom panels the magnetic field on the equatorial plane. The initial (left) and final (right) times are shown. Arrows point to the location of the so-called x-line, the center of reconnection and a few dipolarization fronts (DF) identified by areas of larger vertical field $B_z$.}
\end{figure}
Figure \ref{evolution} shows the evolution of  the system in the equatorial maximum pressure plane. The tail current sheet does not remain perfectly flat as a function of time and develops some warping. We use GSM coordinates ($x$ is \rw{toward the Sun}, $y$ in the dawn to dusk direction and $z$ in the northward direction) and project the information on a flat visualization plane defined for each position $(x,y)$ by the location in $z$ where the trace of the ion pressure tensor \rw{is} maximum.  

The reconnection location can be clearly identified by considering the electron energy flux along $x$ (see arrow in Fig.~\ref{evolution}).  This flux is defined as the third order moment of the distribution function of the velocity of the particles for each species $s$, $\enE_{xs}=\int v_x m_{s}v^{2} f(\bfv)d\bfv /2$. The position of the x-line can be identified as the narrow white band, roughly located in $x \in [-30, -20]$\re between the positive (green) and the negative (cyan) electron energy flux. This is the condition at the x-line where the energy flux is earthward on the Earth side and tailward on the tail side. The inappropriately called  x-line is very far form being the straight line assumed in the common analytical models that impose periodic boundary conditions in the dawn dusk direction $y$. 

In the two reconnection outflow regions, the vertical component of the magnetic field has considerable structure (right panels in Fig.~\ref{evolution}). These are fronts of magnetic energy traveling earthward on the earthward side (\rw{called} dipolarization fronts) or tailward in the tailward side.  

\begin{figure}[htp]
\centering 
\includegraphics[width=\columnwidth]{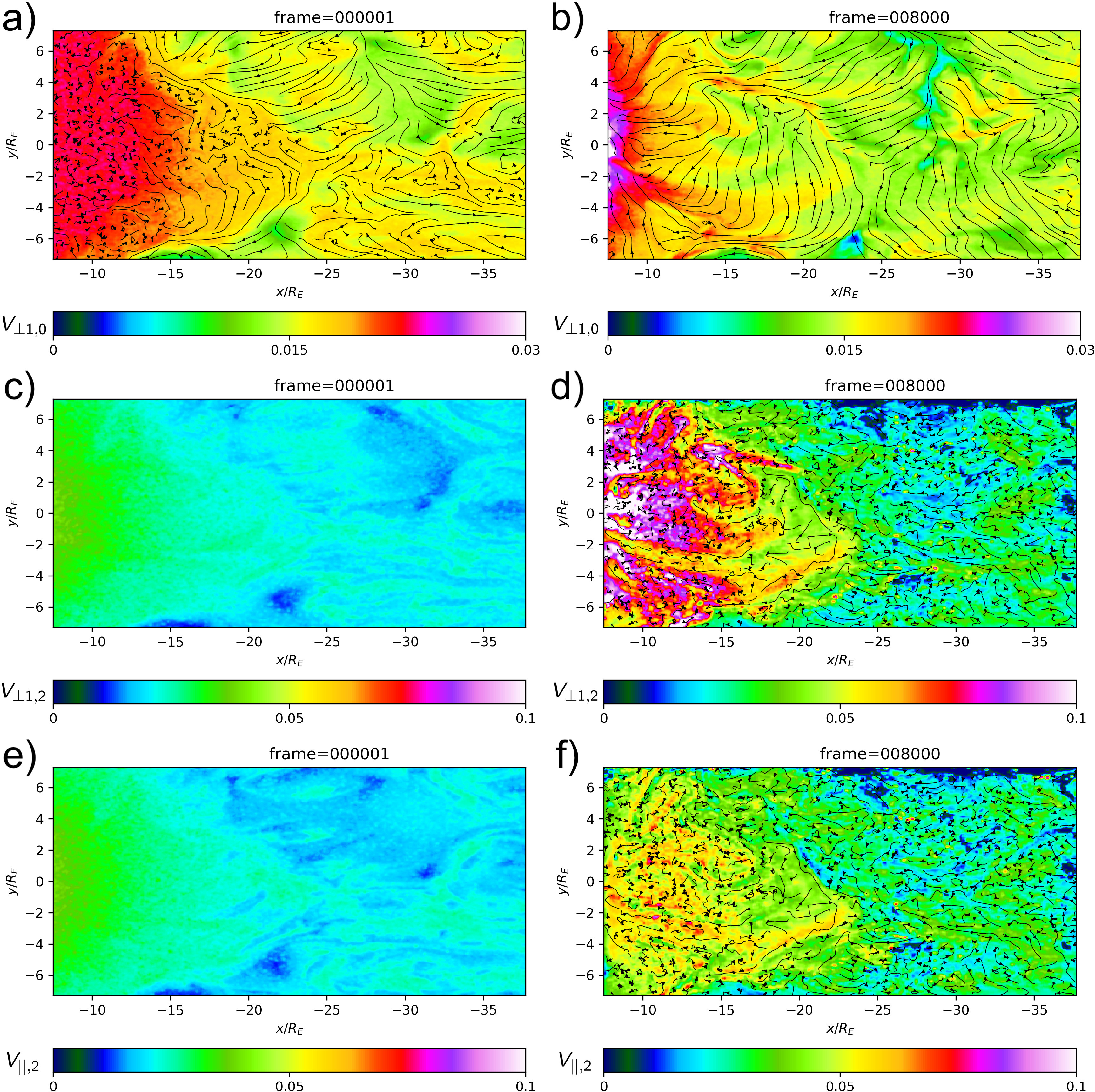}
\caption{ \label{energization} Temporal evolution of the thermal speed (normalised to the speed of light) defined from the second moment of the electron velocity distribution. Two species are shown: the bulk Maxwellian electrons (species 0,top, a-b) and the kappa distributed electrons (species 2 perpendicular in the middle, c-d, and parallel on the bottom, e-f). The black lines show \rw{the} flow field on the equatorial plane for both electron species (initially there is no flow for species 2). The kappa electrons are much more turbulent than the bulk electrons.}
\end{figure}

The temperature of the kappa tail increases by an order of magnitude in vast regions of the \jb{simulation} domain, see Fig.~\ref{energization}. The process of energization is non-uniform. The region near the x-line cools while the heating strongly correlates with the regions of intensified vertical magnetic field (both positive and negative). This effect is especially strong for the kappa tail electrons. One can follow the complicated structure in the earthward region to the left of $x=-25$\re, the same features are \jb{seen} in $B_z$ and in $v_{th,2,\perp}$.

The process of heating for all species is highly \rw{an}isotropic. \jb{However} it remains fairly gyrotropic except for the region of the x-line itself where gyrotropy is broken. Comparing the middle and \gl{bottom} columns of Fig.~\ref{energization}, it is evident \jb{that } the heating is predominant in the perpendicular direction. 
\jb{Overall the} heating is clearly not laminar. 

Next, we investigate the fluctuations of the conversion rate of electromagnetic energy into particle energy, $\bfJ_s \cdot \bfE$ as a function of  the earthward ion flow, $V_{ix}$  and the vertical component of the magnetic field $B_z$. The former  is a proxy for the distance away from the reconnection x-line as the reconnection outflow speed increases away from the reconnection site that is a stagnation point. 


Figure \ref{conditional} shows the results of this conditional fluctuation analysis for the power law electrons, species 2. 
The energization has a wide range of fluctuations at all distances from the x-line (indicated by $V_{ix}$). The turbulence is intermittent. \jb{This is determined both by visual inspection and by noticing the non-Gaussian aspect of the fluctuation spectrum (i.e. non-parabolic in panel c of Fig. \ref{conditional}), which is known to be a good indicator of intermittency~\citep{marsch1994non}.}

Taking the mean and the standard deviation of the distribution of fluctuations, 
we observe that there is a net energization for velocities between 0 and positive 200\rw{k}m/s, the region of the earthward outflow, and for \rw{the} positive vertical component of $B_z$. But this net heating comes as an average of a highly turbulent spectrum of fluctuations where the standard deviation exceeds largely the mean value. 
We conclude that the energization is then caused by the turbulent cascade of energy concentrated in the earthward regions of larger vertical field\jb{s}. This finding is consistent with  adiabatic acceleration~\citep{pan2012adiabatic,sm1, akhavan2019mms,egedal2012large}.


 In summary, we have shown that using a combined MHD and PIC analysis, where one \jb{electron-ion population} of computational particles describes the bulk and a second the high energy tail, we can replicate the observed formation of a high energy power law tail during reconnection in Earth\rw{'s} magnetotail. \jb{In contrast to observations that are  local and instantaneous, the simulation provides us} a global view of the temporal evolution that allows us to investigate the mechanisms of energization \rw{by} using statistical methods. We find that there is a strong energization in the earthward outflow from reconnection and, within it, especially in the \rw{regions} of intensified vertical field. The mechanism of energy transfer that forms the power law tail is characterized by highly fluctuating energy exchanges, with the standard deviation largely exceeding the mean. 
 
 While the study reported here focuses on the conditions observed in  Earth's magnetotail, the mechanisms described are based on the fundamental properties of turbulence and reconnection and it can be expected that similar processes might be present in all instances in laboratory or natural plasmas where reconnection outflows develop \rw{turbulence}.
 
\textit{The work at UCLA was supported by a
NASA grants
80NSSC17K0014 and 80NSSC19K0846, and NSF grant AGS-
1450864. This project has received funding from the European Union's Horizon 2020 research and innovation programme under grant agreement No. 776262 (AIDA). This
research used resources of the National Energy Research
Scientific Computing Center, which is supported by the Office
of Science of the US Department of Energy under Contract No.
DE-AC02-05CH11231. Additional computing has been provided
by NASA NAS and NCCS High Performance Computing,
by the Flemish Supercomputing Center (VSC) and by a
PRACE Tier-0 allocation. 
}\vfill
\newpage
\begin{figure}[htp]
\includegraphics[width=\linewidth]{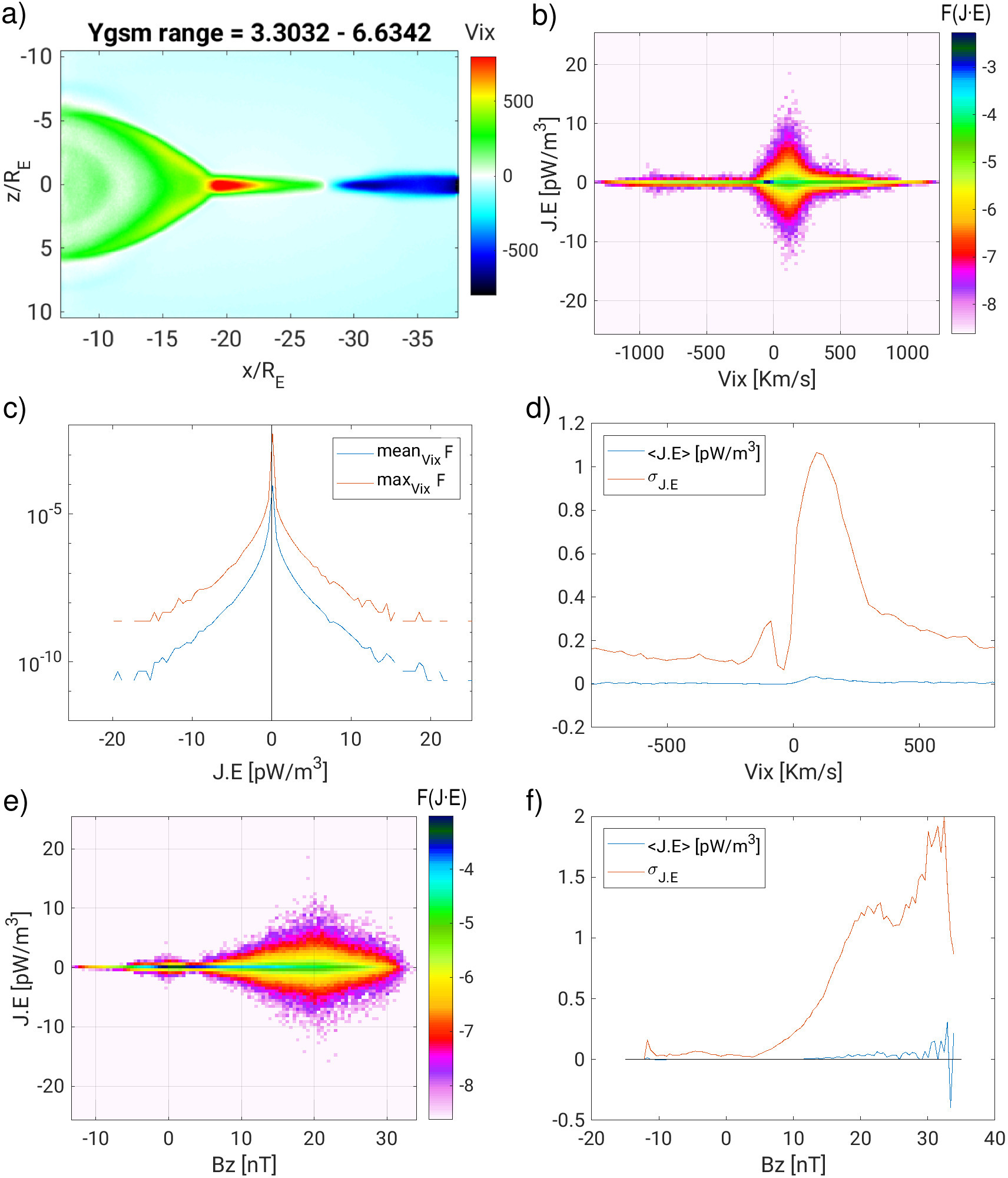}  
\clearpage 
\end{figure}
\captionof{figure}{ Conditional bivariate \rw{analysis of} the fluctuations of the energy exchange for electron species 2 (kappa tail), $\bfJ_2\cdot \bfE$, as a function of $V_{ix}$ and $B_z$ at cycle 1000, corresponding to t=3.75s, considering the range $y=[3.3,6.6]$\re. 
The panels show:
a) False color representation of $V_{ix}$, averaged over $y$ in the range indicated\rw{,}  b) conditional fluctuation  spectrum (measured as $\log_{10}$ of the fraction F of points in the grid) at different $V_{ix}$, integrated over $B_z$, showing the dominance of positive transfer (i.e. towards the particles) over negative (i.e. towards the field), 
c) mean and maximum (along $V_{ix}$) value of the distribution of fluctuations $\bfJ_2\cdot \bfE$ showing that the positive side dominates over the negative side and showing a clear non-parabolic shape, indicative of intermittency\rw{, and}
d) mean and standard deviation of $\bfJ_2\cdot \bfE$ as a function of  $V_{ix}$. This shows clearly  energization  for positive $V_{ix}$ between 0 and 200\rw{k}m/s (regions that can be identified as green in panel a), meaning in the earthward flow turbulence deposits more energy on average than it takes away.
Panel e) gives  conditional fluctuation spectrum of $\bfJ_2\cdot \bfE$ at different $B_{z}$, integrated over $V_{ix}$, indicating that energy exchange is  concentrated in regions of positive normal field, peaking at around 20\si{nT} and
f)  gives mean and standard deviation of $\bfJ_2\cdot \bfE$ as a function of  $B_{z}$, showing that at positive normal fields the energy exchange has a net positive value causing electron energization.}
\label{conditional}

\bibliographystyle{apsrev4-2} 

\appendix

\section{Supplemental Material 1: Alternative initialization for a Maxwellian with importance sampling biased towards the high-energy tail.}

As discussed in the main article text, we carried out a second simulation with just two species, one for the ions and one for the electrons. However instead of using the Box-Müller algorithm to generate particles from a Maxwellian, we used importance sampling. We sampled particles in a uniform distribution in velocity between $-5 V_{th} $and $+5 V_{th}$ , assigned to each of the sampled particles the weight required by the importance sampling algorithm [1] and added the local drift speed. The method is similar to that proposed by Ref. [2] but differs because we still sample the initial speed randomly from a uniform distribution rather than generating it on a uniform velocity grid. Also, positions are randomly selected within a cell. This randomness avoids the spurious multi-beam numerical instability [3].

\begin{figure}[htp]
\centering 
\includegraphics[width=\columnwidth]{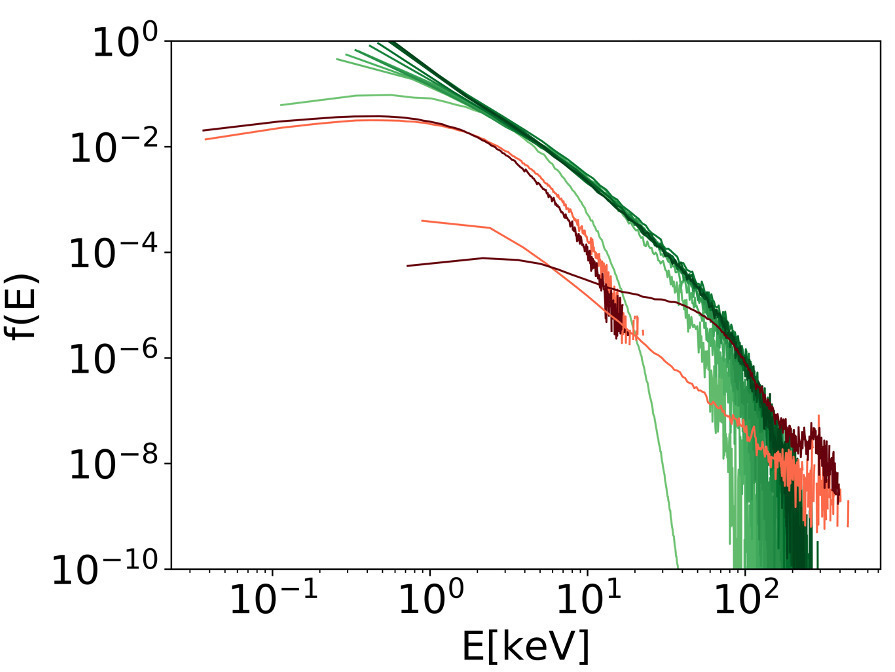}
\caption{ \label{suppl1} Velocity distribution function at t=0, 0.75, 1.5, 2.25, 3.0, 3.75, 7.5, 11.25,15.0 and 18.75 seconds from the start of the run is shown in progressively darker shades of purple. The initial (light red) and final (dark red) distributions for the run with 2 populations per species of the main manuscript are reported for comparison.  The particles reported are in the physical domain: $x/ R_E=[-38,-7]$, $y/ R_E =[5.3,5.6] $, $z/ R_E =[-.2,.2]$.}
\end{figure}

Otherwise, the simulation is identical. Figure \ref{suppl1} shows the evolution of the electron velocity distribution. As can be seen, the initial Maxwellian rapidly evolves towards a power law distribution and the final state is similar to that reported in the main manuscript and based on the 2 initial populations per species, a core and a tail. 

The results shown in Fig. \ref{suppl1} are presented using logarithmic axes and the distribution is normalized so that the integral of the distribution is unitary in both simulations.  We learned two things from this analysis. First, this test validates the approach used in the main manuscript, showing that the details of initialization do not matter in enabling the mechanism of particle acceleration that leads to the power law. While common mechanisms of numerical heating intrinsic to PIC might be at work in both cases, the initialization does not have a key effect: despite the two methods being radically different, the results are similar. Second, as noted in the main manuscript, the formation of a power law tail does not require seeding a power tail at time t=0. In the importance sampling initialization, the velocity distribution is Maxwellian at the initial time but a power law forms quickly.

We can obtain more insight into the mechanism of acceleration from Fig. S2, reporting the statistical distribution of electrons according to the local vertical component of the magnetic field and their perpendicular energy. As can be seen, the particle energy increases with the local magnetic field while the first adiabatic invariant ($\mu=mv_\perp^2/2B$) remains relatively constant. This behavior is consistent with adiabatic acceleration, i.e. the energy increases where B increases to keep the first adiabatic invariant constant.

 \begin{figure}[htp]
\centering 
\includegraphics[width=\columnwidth]{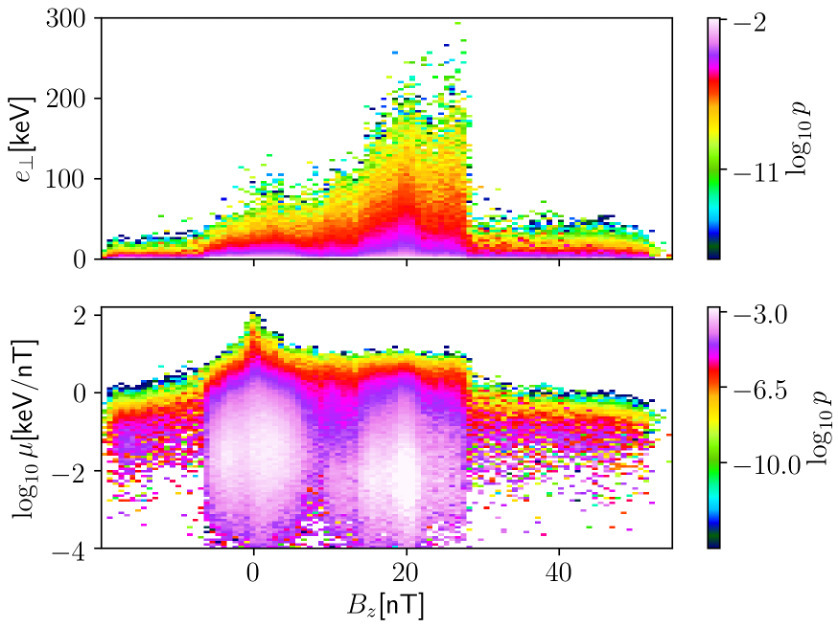}
\caption{ \label{suppl2} probability distribution of the electrons as a function of Bz and their perpendicular energy (top) and their first adiabatic invariant $\mu=mv_\perp^2/2B$ (bottom). High negative and positive values of Bz correspond to regions not affected by reconnection, visible as a sharp drops in the probability. The particles reported are in the physical domain: $x/ R_E =[-38,-7], y/ R_E =[5.3,5.6] ], z/ R_E =[-.2,.2]$.}
\end{figure}

The results in Fig. \ref{suppl2} present two trends. First, there is a general increase in the probability at a given perpendicular energy as a function of Bz. This seems to indicate adiabatic acceleration as the electrons convect earthward (overall increase in Bz).  Second, there are two clear drops (at -5nT and +28nT in Bz) where the probability drops dramatically. These more extreme vertical magnetic fields (large and negative in the tailward direction, large and positive in the Earthward direction) correspond to regions of the domain not affected by reconnection-induced acceleration.

While the presence of acceleration is common in the run shown here and in that shown in the main document, there are interesting practical differences. 

The importance sampling is natural and agnostic. The simulation of the main manuscript has an initial power law tail arbitrarily chosen to expand by 3 orders of magnitude the dynamical range of the distribution followed.  The choice of three orders of magnitude has no particular reason and is arbitrary. We tried also 2 orders and the results were not qualitatively different.  However, importance sampling has no free parameter and no arbitrariness (except fixing the initial range at $\pm5V_{th}$ but we tried also $\pm3V_{th}$ and the difference is negligible). Importance sampling covers the range completely without any gaps.

On the other hand, the two-population approach does provide a larger dynamical range, expanding the coverage at both the low and high energy range. However, there is a gap in between the high energy and low energy populations that importance sampling does not show.

In summary, there is no immediate way to prefer one solution over the other. Both provide a viable way to investigate the high energy spectrum under sampled in the Box-Müller algorithm to generate particles from a Maxwellian.

[1] Kroese, D. P., \& Rubinstein, R. Y. (2012). Monte carlo methods. Wiley Interdisciplinary Reviews: Computational Statistics, 4(1), 48-58.

[2] J. Byers, Noise Suppression Techniques in Macroparticle Models of Collisionless Plasmas., Tech. Rep. (California Univ., Livermore. Lawrence Radiation Lab., 1970).

[3] C. Birdsall and A. Langdon, Plasma Physics Via Computer Simulation (Taylor \& Francis, London, 2004). See Chapter 14.

[4] Pan, Q., Ashour‐Abdalla, M., El‐Alaoui, M., Walker, R. J., \& Goldstein, M. L. (2012). Adiabatic acceleration of suprathermal electrons associated with dipolarization fronts. Journal of Geophysical Research: Space Physics, 117(A12).

\section{Supplemental Material 2: Movie.}
A movie of the 3D volume rendering of the ion current at the end of the simulation (cycle 8000) with magnetic field lines selected progressively closer to the viewer. File provided.

\end{document}